%% file: Cross.tex
\documentclass[sigconf]{acmart}

\usepackage{booktabs} 
\usepackage{threeparttable}
\usepackage{colortbl} 
\usepackage{color}
\usepackage{textcomp}
\usepackage{soul} 
\usepackage{amsmath}
\usepackage{comment}
\usepackage{multirow}
\usepackage{algorithm} 
\usepackage{algpseudocode} 
\usepackage{bbding} 

\newtheorem{myDef}{Definition}

\copyrightyear{2018} 
\acmYear{2018} 
\setcopyright{acmcopyright}
\acmConference[SIGIR '18]{The 41st International ACM SIGIR Conference on Research \& Development in Information Retrieval}{July 8--12, 2018}{Ann Arbor, MI, USA}
\acmBooktitle{SIGIR '18: The 41st International ACM SIGIR Conference on Research \& Development in Information Retrieval, July 8--12, 2018, Ann Arbor, MI, USA}
\acmPrice{15.00}
\acmDOI{10.1145/3209978.3210032}
\acmISBN{978-1-4503-5657-2/18/07}

\begin{document}
\fancyhead{}
\title[Cross-language Citation Recommendation via HRLHG]{Cross-language Citation Recommendation via Hierarchical Representation Learning on Heterogeneous Graph}

\author{Zhuoren Jiang}
\affiliation{
\institution{$^1$Sun Yat-sen University}
\city{Guangzhou} 
\country{China}}
\affiliation{\institution{$^2$Peking University, Beijing, China}}
\email{jiangzhr3@mail.sysu.edu.cn}

\author{Yue Yin}
\affiliation{
\institution{Beijing Normal University}
\city{Beijing} 
\country{China}}
\email{bnuyinyue@outlook.com}

\author{Liangcai Gao}
\authornote{Corresponding author}
\affiliation{
\institution{Peking University}
\city{Beijing} 
\country{China}}
\email{glc@pku.edu.cn}

\author{Yao Lu}
\affiliation{
\institution{Sun Yat-sen University}
\city{Guangzhou} 
\country{China}}
\email{luyao23@mail.sysu.edu.cn}

\author{Xiaozhong Liu}
\authornotemark[1]
\affiliation{
\institution{$^1$Alibaba Group}
\city{Seattle \& Hangzhou} 
\country{China}}
\affiliation{
\institution{$^2$Indiana University Bloomington}
\city{Bloomington} 
\state{IN}
\country{USA}}
\email{liu237@indiana.edu}

\begin{abstract}
While the volume of scholarly publications has increased at a frenetic pace, accessing and consuming the useful candidate papers, in very large digital libraries, is becoming an essential and challenging task for scholars. Unfortunately, because of language barrier, some scientists (especially the junior ones or graduate students who do not master other languages) cannot efficiently locate the publications hosted in a foreign language repository. In this study, we propose a novel solution, cross-language citation recommendation via Hierarchical Representation Learning on Heterogeneous Graph (HRLHG), to address this new problem. HRLHG can learn a representation function by mapping the publications, from multilingual repositories, to a low-dimensional joint embedding space from various kinds of vertexes and relations on a heterogeneous graph. By leveraging both global (task specific) plus local (task independent) information as well as a novel supervised hierarchical random walk algorithm, the proposed method can optimize the publication representations by maximizing the likelihood of locating the important cross-language neighborhoods on the graph. Experiment results show that the proposed method can not only outperform state-of-the-art baseline models, but also improve the interpretability of the representation model for cross-language citation recommendation task.
\end{abstract}

\settopmatter{printccs=false}

%
%




\maketitle

\input{intro}
\input{problem}
\input{method}

\input{experiment}
\input{review}
\input{conclusion}
\begin{acks}
The work is supported by the National Science Foundation of China (11401601, 61573028, 61472014), Guangdong Province Frontier and Key Technology Innovative Grant (2015B010110003, 2016B030307003), Health \& Medical Collaborative Innovation Project of Guangzhou City, China (201604020003) and the Opening Project of State Key Laboratory of Digital Publishing Technology.
\end{acks}
\bibliographystyle{ACM-Reference-Format}
\bibliography{refs} 

\end{document}

%% file: intro.tex
\vspace{-2ex}\section{Introduction}

\textit{``It takes me a lot more time to find a useful paper... and it takes me even longer to read it... ''} while a non-English speaking PhD student complained this in a seminar, other PhD candidates, in the similar background, agreed with her and they shared the same frustration when they are trying to find and consume the helpful English publications. Professor's (a native speaker) response came later as a relief \textit{``well, I agree, but my problem is even bigger... I cannot read the papers in your language at all...''} This dialog initialized our thinking about this new problem - \textbf{Cross Language Publication (Citation) Recommendation}, a.k.a. how can we propose a useful method/system to assist scholars to efficiently locate the useful publications written in different languages (a typical scenario of this task is to help non-English speaking students to search for useful English papers). Increased academic globalization is forcing a scholar to break the linguistic boundaries, and English (or any other dominant language) may not always serve as the gatekeeper to scientific discourse. 

Unfortunately, existing academic search engines (e.g. Google Scholar, Microsoft Academic Search, etc.) along with many sophisticated retrieval and recommendation algorithms \cite{he2010context, jiang2015chronological,tang2009discriminative} cannot cope with this problem efficiently. For instance, most of the existing citation recommendation algorithms work in a monolingual context, and the scholarly graph-based random walk may not work well in a multilingual environment (section 4 will prove this). 

Moreover, Cross-language Citation Recommendation (CCR) can be a quite challenging problem comparing with classical scholarly recommendation due to the following reasons:

\textbf{Information need shifting}. Different from monolingual citation recommendation, we cannot directly calculate the relevance between the papers written in two different languages. A straightforward solution is to utilize machine translation (MT) \cite{bahdanau2014neural} to translate the query content (e.g., keyword, text or user profile), then, use existing matching models \cite{zhai2001study, guo2016deep} to recommend the proper papers in target language. However, MT based methods and the CCR task can be fundamentally different. The goal of MT is to find a target text given a source text based on the same semantic meaning \cite{bahdanau2014neural} (e.g., find the papers contain exact or similar matched phrases or sentences), while the CCR task is focusing on recommending ``relevant'' papers in target language to the given query in the source language \cite{tang2014cross}. When research context changes, content translation may not perform well. For instance, in Chinese/Japanese research context, machine learning methods can be important for word segmentation studies, which may not be the case for the English counterpart. MT approach cannot address this kind of information need shifting problem. 

\textbf{Sparse inter-repositories citation relations}. Besides textual information, citation relations are quite important for citation recommendation. In the prior studies, recommendation algorithms can learn the ``relevance'' by using citation relations on a graph \cite{ren2014cluscite, liu2014meta}. However, compared to the enormous monolingual citation relations, cross-language citations can be very sparse. For instance, in a computer science related bilingual (Chinese-English) context, we find the papers in ACM and Wanfang\footnote{One of the biggest digital libraries in Chinese.}, on average, have about 28 times more monolingual citation relations than cross-language ones. It is difficult to effectively employ the citation relations for cross-language citation recommendation by using classical graph mining methods.

\textbf{Heterogeneous information environment}. Intuitively, one could integrate the cross-language content semantics, citation relations and other useful heterogeneous information (e.g., keywords and authors) to address CCR. However, most existing text or graph based ranking algorithms rely on a set of human defined rules (e.g., sequential relation path \cite{lao} and meta-path \cite{sun2011pathsim}) to integrate different kinds of information. On a complex cross-language scholarly graph, this kind of handcrafting features can be time-consuming, incomplete and biased.

\begin{figure}[htbp]\centering
 	\includegraphics[width=1.0\columnwidth]{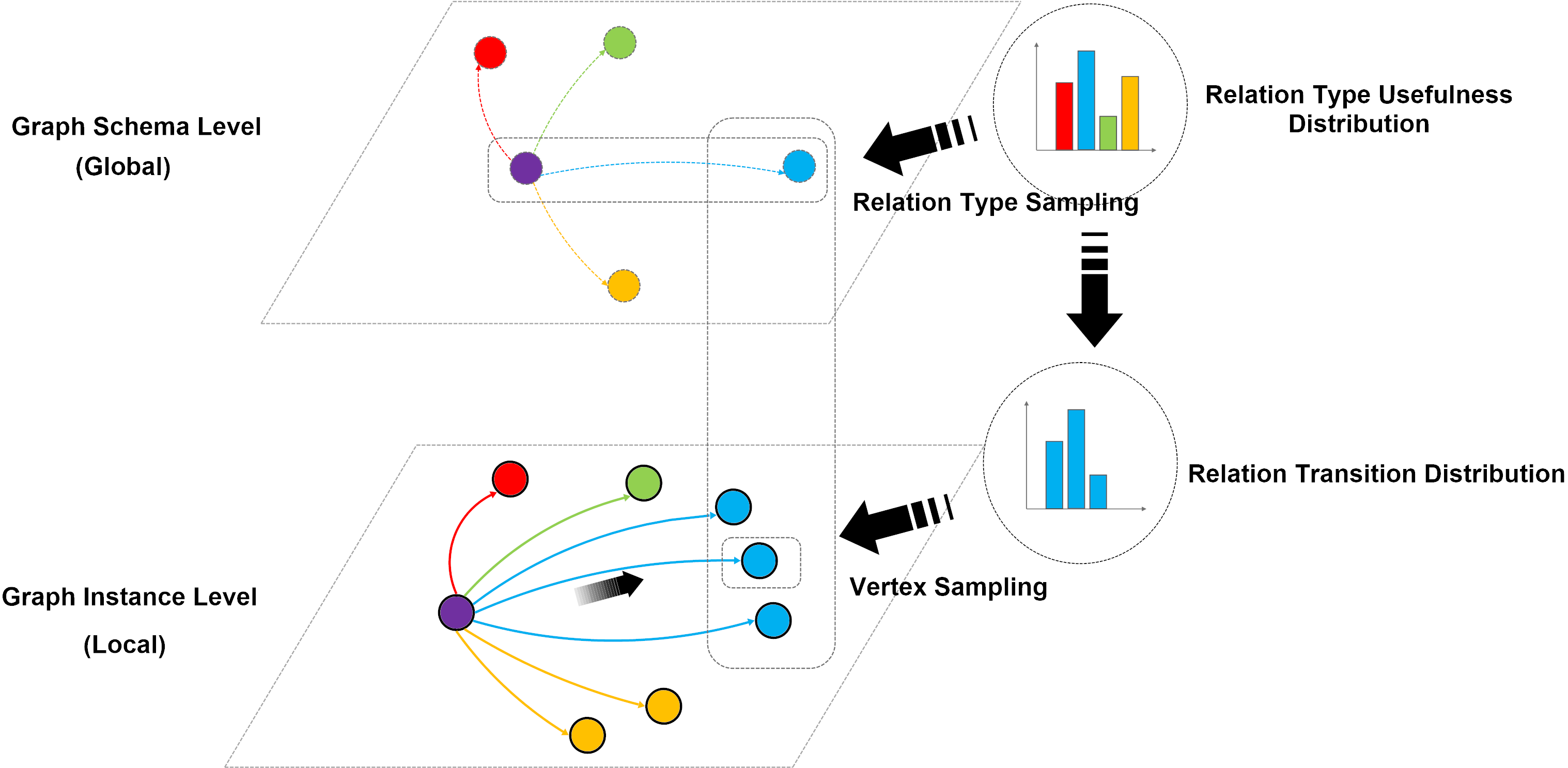}
 	\caption{Hierarchical random walk illustration (different colours denote different types)}
 	\label{fig:example}
\end{figure}

To address these challenges, in this study, we propose a novel solution, \textbf{Hierarchical Representation Learning on Heterogeneous Graph (HRLHG)}, for cross-language citation recommendation. By constructing a novel cross-language heterogeneous graph with various types of vertexes and relations, we ``semantically'' enrich the basic citation structure to carry more rich information. To avoid the handcrafting feature usage, we propose an innovative algorithm to project a vertex (on the heterogeneous graph) to a low-dimensional joint embedding space. Unlike prior hyperedge or meta-path approaches, the proposed algorithm can generate a set of Relation Type Usefulness Distributions (RTUD), which enables fully automatic heterogeneous graph navigation. As Figure \ref{fig:example} shows, the hierarchical random walk algorithm enables a two-level random walk guided by two different sets of distributions. The global one (relation type usefulness distributions) is designed for graph schema level navigation (task-specific); while the local one (relation transition distributions) targets on graph instance level walking (task-independent). 

By using HRLHG, we can recommend a list of ranked cross-language citations for a given paper/query in the source language. We evaluate the proposed algorithm in Chinese and English scholarly corpora, i.e., Wanfang and ACM digital libraries. The results demonstrate that the proposed approach is superior than state-of-the-art models for cross-language citation recommendation task.

\textbf{The contribution of this paper is fourfold.} First,
we propose a novel method (Hierarchical Representation Learning on Heterogeneous Graph) to characterize both the global and local semantic plus topological information for the publication representations. Second, we improve the interpretability of the publication representation model. By using an iterative EM (expectation-maximization) approach, the proposed algorithm can learn the implicit biases for cross-language citation recommendation, which significantly differs from classical heterogeneous graph mining algorithms. Third, we apply the proposed embedding method for a novel cross-language citation recommendation task. An experiment on real-world bilingual scientific datasets is employed to validate the proposed approach. Last but not least, although in this study we focus on cross-language citation recommendation task, the proposed method can be generalized for different tasks that are based on heterogeneous graph embedding learning.

%% file: problem.tex
\vspace{-1ex}\section{Problem Formulation} \label{sec:problem}

Compared to the homogeneous graph, the heterogeneous graph has been demonstrated as a more efficient way to model real world data for many applications, it represents an abstraction of the real world, focusing on the objects and the interactions between the objects \cite{liu2014meta}. Formally, following the works \cite{sun2011pathsim, dong2017metapath2vec}, we present the definitions of a heterogeneous graph with its schema.

\begin{myDef}
\textbf{Heterogeneous Graph}, namely heterogeneous information network, is defined as a graph $G=(V,E,\tau, \gamma )$, where $V$ denotes the vertex set, and $E\subseteq  V\times V$ denotes the edge (relation) set. $\tau$ is the vertex type mapping function, $\tau : V \rightarrow \mathbb{N}$ and $\mathbb{N}$ denotes the set of vertex types. $\gamma$ is relation type mapping function, $\gamma : E \rightarrow \mathbb{Z}$ and $\mathbb{Z}$ denotes the set of relation types. $\left | \mathbb{N} \right | + \left | \mathbb{Z} \right | > 2 $.
\end{myDef}

\begin{myDef}
\textbf{Graph Schema}. The graph schema is a meta template for a heterogeneous graph $G=(V,E,\tau, \gamma)$, denoted as $S_{G}=(\mathbb{N},\mathbb{Z})$.
\end{myDef}

The graph schema is used to specify type constraints on the sets of vertexes and relations of a heterogeneous graph. A graph that follows a graph schema is then called a \textbf{Graph Instance} following the target schema \cite{liu2014meta}.

\begin{myDef}
\textbf{Cross-language citation recommendation}. The CCR problem can be defined as a conditional probability  $Pr(p_{c}|p_{q})$, i.e., the probability of $p_{c}$ in target language given a particular query paper $p_{q}$ in the source language: $$Pr(p_{c}|p_{q}) = \Delta \left ( \phi (p_{q}),\phi (p_{c}) \right )$$ where $\phi$ is a representation function, which can project each paper to a low-dimensional embedding space. $\Delta$ is probability scoring function based on the learned publication embeddings. 
\end{myDef}

The CCR problem can be formalized as:

\begin{itemize}
\item \textbf{Input}: A query paper (or partial text/keywords in the query paper) in a source language. 
\item \textbf{Output}: A list of ranked papers in target language that could be potentially cited or useful given the input paper. 
\end{itemize}

In this study, we investigate the novel method to enhance the representation learning function $\phi$ for CCR. More detailed method will be introduced in Section \ref{sec:method}.

%% file: method.tex
\vspace{-1ex}\section{Hierarchical Representation Learning on Heterogeneous Graph}\label{sec:method}
In this section, we discuss the proposed method in detail.  We first formulate the heterogeneous graph based representation learning framework for CCR task~(\ref{ssec:frame}), then, introduce the hierarchical random walk-based strategy by leveraging the critical relation type usefulness distribution training algorithm~(\ref{ssec:hierarchical})

\vspace{-1ex}\subsection{Heterogeneous Graph Representation Learning Framework for CCR}\label{ssec:frame}
\begin{figure}[htbp]\centering
 	\includegraphics[width=0.9\columnwidth]{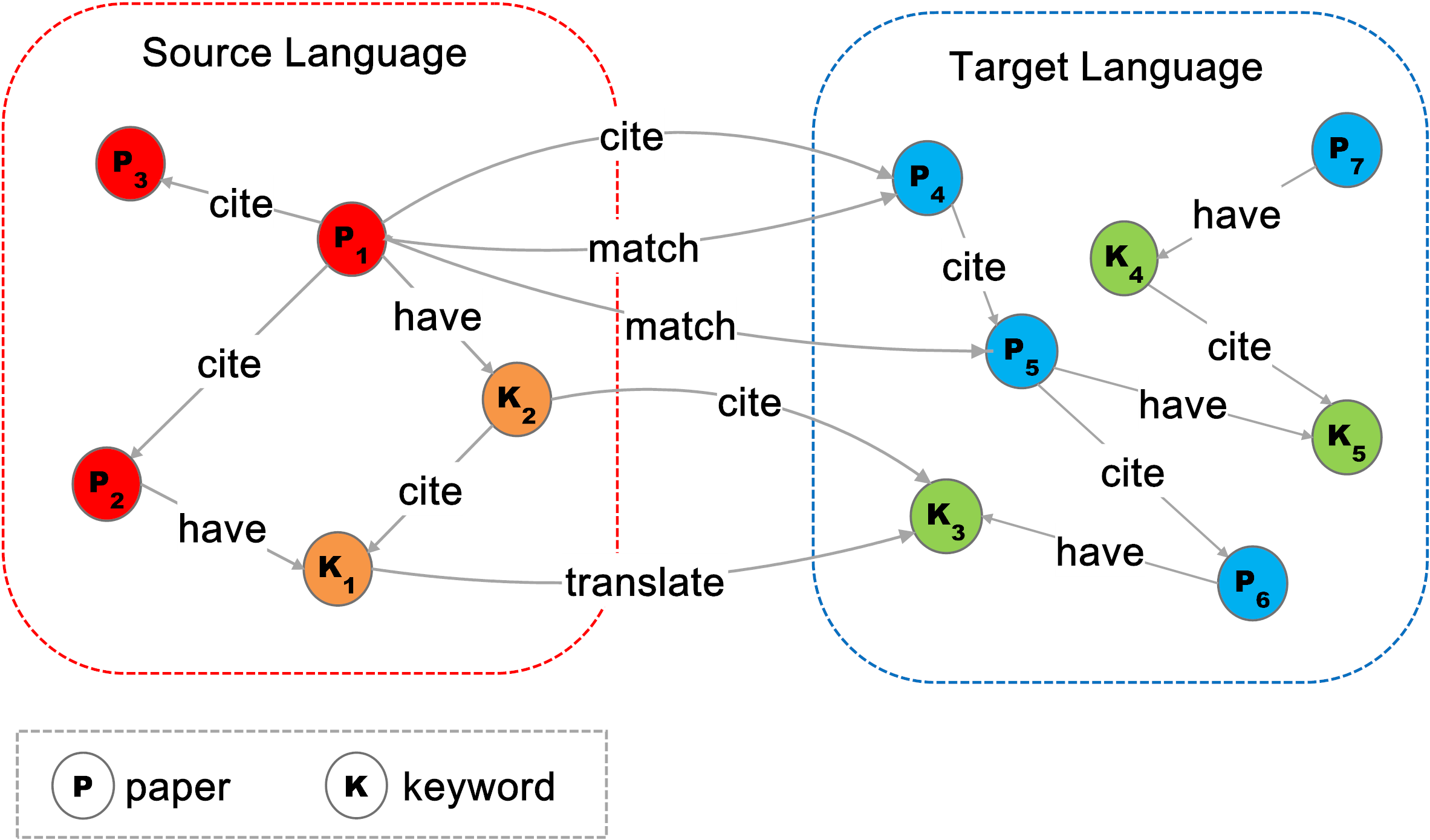}
 	\caption{The constructed cross-language heterogeneous graph (different colours denote different types)}
 	\label{fig:graph}
\end{figure}

Due to the aforementioned challenges of CCR task, the proposed representation model can hardly depend only on textual or citation information. In this study, we integrate various kinds of entities and relations into a heterogeneous graph (as Figure \ref{fig:graph} shows, and the detailed node and type information can refer to Table \ref{tab:graph}). Then, the goal is to design a novel representation learning model to encapsulate both semantic and topological information into a low-dimensional joint embedding for CCR task. 

\vspace{-1ex}

\begin{table}[htb]
\footnotesize
\centering
\caption{Vertexes and relations of cross-language heterogeneous graph}\vspace{-2ex}
\label{tab:graph}
\begin{threeparttable}
\begin{tabular}{l l  p{5.5cm}}
\toprule
\textbf{No.}&\textbf{Vertex} & \textbf{Description}             \\ \midrule
1&$P_{s}$           &  Paper in Source Language       \\ 
2&$P_{t}$           &  Paper in Target Language       \\
3&$K_{s}$           &  Keyword in Source Language     \\
4&$K_{t}$          &   Keyword in Target Language   \\ \midrule
\textbf{No.}&\textbf{Relation}       & \textbf{Description$^\S$}     \\ \midrule
1&$P_{s} \overset{s}{\rightarrow} P_{t}$           &  A paper (in source language) is semantically related to another paper (in target language). We use machine translation\textbf{*} and language model (with Dirichlet smoothing) to generate this relation \cite{zhai2001study}. \\ 
2&$P_{s} \overset{c}{\rightarrow} P_{s}$           &  A Paper (in source language) has a  monolingual citation relation to another paper (in source language).      \\ 
3&$P_{t} \overset{c}{\rightarrow} P_{t}$           &  A Paper (in target language) has a  monolingual citation relation to another paper (in target language).       \\
4&$P_{s} \overset{c}{\rightarrow} P_{t}$           &  A Paper (in source language) has a  cross-language citation relation to another paper (in target language).   \\
5&$P_{s} \overset{h}{\rightarrow} K_{s}$           &  A Paper (in source language) has a keyword (in source language). \\
6&$P_{t} \overset{h}{\rightarrow} K_{t}$           &  A Paper (in target language) has a keyword (in target language). \\
7&$K_{s} \overset{c}{\rightarrow} K_{s}$           &  A keyword (in source language) has a monolingual citation relation to another keyword (in source language)$^\ddagger$. \\
8&$K_{t} \overset{c}{\rightarrow} K_{t}$           &  A keyword (in target language) has a monolingual citation relation to another keyword (in target language)$^\ddagger$.  \\
9&$K_{s} \overset{c}{\rightarrow} K_{t}$           &  A keyword (in source language) has a cross-language citation relation to another keyword (in target language)$^\ddagger$. \\
10&$K_{s} \overset{t}{\rightarrow} K_{t}$       &  A keyword (in source language) is translated into the corresponding keyword (in target language)\textbf{*}. \\
\bottomrule 
\end{tabular}
\begin{tablenotes}
    \footnotesize
    \item *As this study is not focusing on machine translation, we use Google machine translation API (https://cloud.google.com/translate) to translate the paper abstract and keywords.
    \item $^\ddagger$The keyword citation relations are derived from paper citation relations.
    \item $^\S$Because of the space limitation, the detailed relation transition probability calculation can be found at https://github.com/GraphEmbedding/HRLHG.
\end{tablenotes}
\end{threeparttable}
\end{table}\vspace{-1ex}

Formally, given a heterogeneous graph $G=(V,E,\tau, \gamma )$, $\tau : V \rightarrow \mathbb{N}$ is the vertex type mapping function and $\gamma : E \rightarrow \mathbb{Z}$ is the relation type mapping function. The goal of vertex representation learning is to obtain the latent vertex representations by mapping vertexes into a low-dimensional space $\mathbb{R}^{d}$, $d \ll |V|$. The learned representations are able to preserve the information in $G$. We use $f:V\rightarrow \mathbb{R}^{d}$ as the mapping function from multi-typed vertexes to feature representations. Here, $d$ is a parameter specifying the number of dimensions. $f$ is a matrix of size $\left | V \right | \times d$ parameters. The following objective function should be optimized for heterogeneous graph representation learning. 
\begin{equation}\label{equ:max}
\underset{f}{max}\sum_{v\in V} \sum_{n \in \mathbb{N}} \sum_{v_{n}^{c} \in N_{n}(v)} log Pr(v_{n}^{c}|\overrightarrow{f(v)})
\end{equation}
where $N_{n}(v)$ denotes $v$'s network neighborhood (``context'') with the $n^{th}$ type of vertexes. The feature learning methods are based on the Skip-gram architecture \cite{mikolov2013efficient,mikolov2013distributed,bengio2013representation}, which is originally developed for natural language processing and word embedding. Unlike the linear nature of text, the structural and semantic characteristics of graph allow the vertex's network neighborhood, $N(v)$, to be defined in various of ways, i.e., direct (one-hop) neighbors of $v$. It is critical to model vertex neighborhood in graph representation learning. Following the previous network embedding models \cite{perozzi2014deepwalk, grover2016node2vec, dong2017metapath2vec}, in this study, we leverage a random walk-based strategy for every vertex $v \in V$ to generate $N(v)$. For instance, in
Figure \ref{fig:graph}, we can sample a random walk sequence
$\left \{ p_{1}, k_{2},k_{3},p_{6}\right \}$ of length $l =4$, which
results in $N(p_{1}) = \left \{ k_{2} \right \}$,
$N(k_{2}) = \left \{ p_{1},k_{3} \right \}$, $N(k_{3}) = \left \{ k_{2},p_{6}\right \}$ and $N(p_{6}) = \left \{ k_{3}\right \}$ (window size is 1). The detailed description of this method will be introduced in section \ref{ssec:hierarchical}.

$Pr(v_{n}^{c}|\overrightarrow{f(v)})$ defines the conditional probability of having a context vertex $v_{n}^{c} \in N_{n}(v)$ given the node $v$'s representation, which is commonly modeled as a softmax function :
\begin{equation}\label{equ:softmax}
Pr(v_{n}^{c}|\overrightarrow{f(v)}) = \frac{exp(\overrightarrow{f(v_{n}^{c})}\cdot \overrightarrow{f(v)})}{\sum_{u\in V}exp(\overrightarrow{f(u)}\cdot \overrightarrow{f(v)})}
\end{equation}

In this study, we use a \textbf{Heterogeneous Softmax} function for conditional probability $Pr(N(v)|\overrightarrow{f(v)})$ calculation \cite{dong2017metapath2vec}:
\begin{equation}\label{equ:hsoftmax}
Pr(v_{n}^{c}|\overrightarrow{f(v)}) = \frac{exp(\overrightarrow{f(v_{n}^{c})}\cdot \overrightarrow{f(v)})}{\sum_{u_{n}\in V_{n}}exp(\overrightarrow{f(u_{n})}\cdot \overrightarrow{f(v)})}
\end{equation}
where $V_{n}$ is the vertex set of type $n$ in $G$. Different from common Skip-gram form, the \textbf{Heterogeneous Skip-gram} with heterogeneous softmax function can specify one set of multinomial distributions for each type of neighborhood in the output layer of the Skip-gram model. Stochastic gradient ascent is used for optimizing the model parameters of $\overrightarrow{f}$. Negative sampling \cite{mikolov2013distributed} is applied for optimization efficiency. Especially, for ``heterogeneous softmax'', the negative vertexes are sampled from the graph according to their type information \cite{dong2017metapath2vec}. 

Recall the CCR definition in section \ref{sec:problem}, given a query paper $p_{q}$ in source language, the problem is to compute the recommendation probability $Pr(p_{c}|p_{q})$ of a candidate paper $p_{c}$ in target language, with representation function $\phi$ and probability scoring function $\Delta$. In this study, $\phi$ is the optimized heterogeneous vertex representation $\overrightarrow{f}$, and we use cosine similarity with Relu function for $\Delta$:
\begin{equation}\label{equ:hsoftmax}
Pr(p_{c}|p_{q}) = Max(0,  \frac{\overrightarrow{f(p_{q})} \cdot \overrightarrow{f(p_{c})}}{\left \| \overrightarrow{f(p_{q})}  \right \| \left \| \overrightarrow{f(p_{c})}  \right \|})
\end{equation}

\vspace{-2ex}\subsection{Hierarchical Representation Learning on Heterogeneous Graph}\label{ssec:hierarchical}
In this section, we propose a novel hierarchical random walk-based strategy for vertex neighborhoods on heterogeneous graph. Before moving on, let's clarify four challenges for random walk-based graph embedding models.

(1) The existing homogeneous random walk-based embedding approaches, e.g., \cite{perozzi2014deepwalk,grover2016node2vec}, cannot be directly applied to address the heterogeneous graph problems. For instance, as Figure \ref{fig:graph} shows, between the paper pair $p_{1}$ and $p_{3}$, there are two different types of relations: $p_{1} \overset{s}{\rightarrow} p_{3}$ ($p_{1}$ is semantically related to $p_{3}$) and $p_{1} \overset{c}{\rightarrow} p_{3}$ ($p_{1}$ cites $p_{3}$), but the homogeneous random walk-based approaches cannot distinguish the difference of relation types, then the neighborhood generating could be problematic for further representation learning. 

(2) Recent heterogeneous graph embedding algorithms \cite{dong2017metapath2vec,fu2017hin2vec} require a domain expert to generate random walk hypotheses, which can be inconvenient and problematic for complex heterogeneous graphs. 

(3) Insufficient global information. For each step in the walk, a lot of random walk-based models are solely depending on the (local) network topology of the vertexes, but global information, i.e., graph schema information, may bring important information to navigate the walker on the graph.

(4) Most existing graph embedding methods aim to encode the topological information of the graphs, which are task independent. For instance, as described in Table \ref{tab:graph}, the vertexes and relations with transition probability are fixed after the graph is constructed. We argue that, the learned representation should be optimized for different tasks, e.g., cross-language citation recommendation task for this study. A flexible representation mechanism can be important to address recommendation problem via heterogeneous graph. For instance, on a complex graph, some kinds of relations can be more important for random walk than others given the task (conditional relation type usefulness probability given the task). 

\begin{figure}[htbp]\centering
 	\includegraphics[width=0.8\columnwidth]{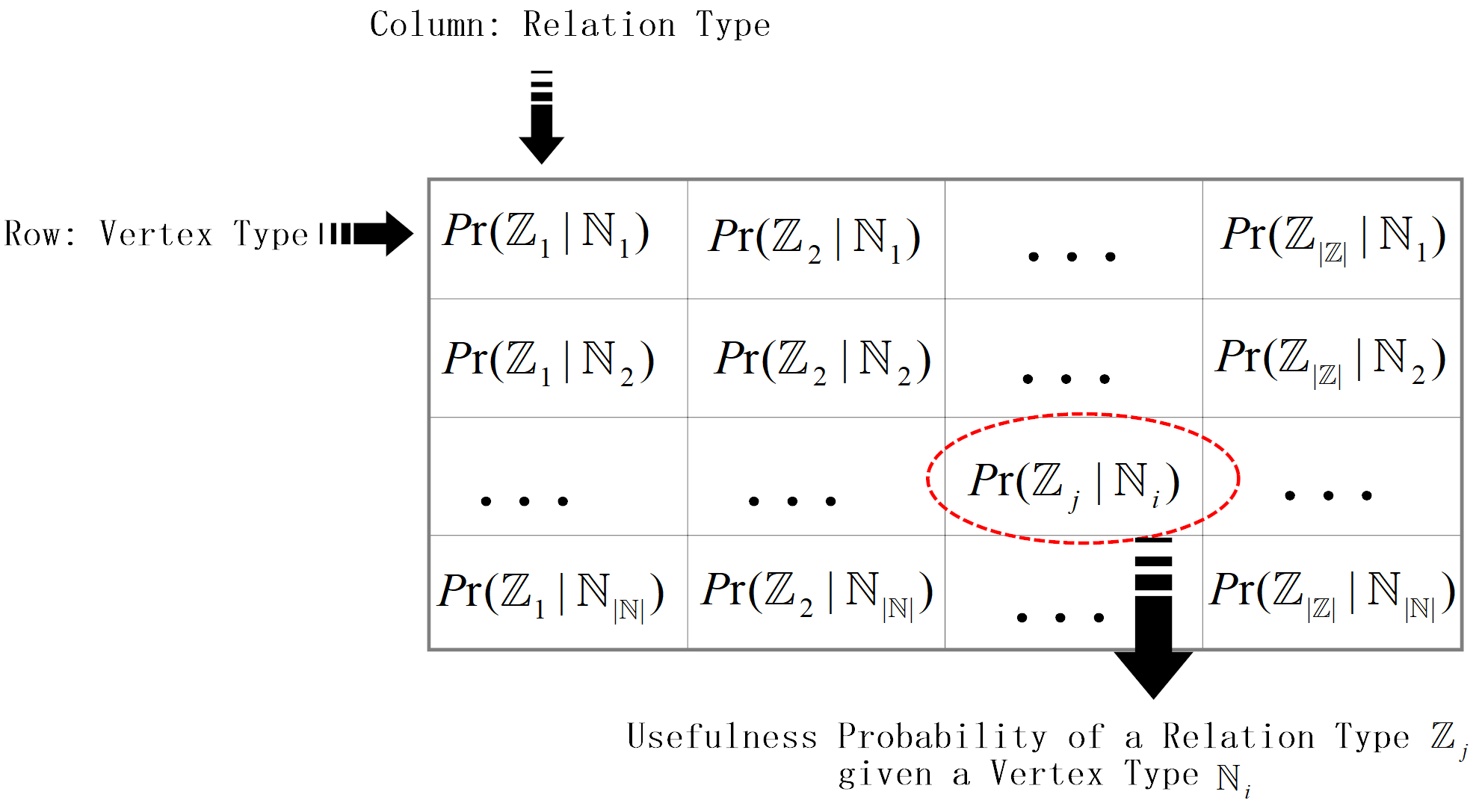}\vspace{-2ex}
 	\caption{Relation type usefulness distributions illustration}
 	\label{fig:RTUD}
\end{figure}

To address these challenges, we propose a Hierarchical Representation Learning on Heterogeneous Graph (HRLHG) method. By introducing a set of Relation Type Usefulness Distributions (RTUD) on graph schema, the hierarchical (two-level) random walk algorithm can be more appropriate for heterogeneous network structure. As RTUD can be automatically learned, we don't need expert knowledge (e.g., generating meta-path) for representation learning. Meanwhile, by using RTUD, we not only bring global information for guiding the random walk, but also utilize the task specific information for optimizing the random walk generation.

Given a specific task $T$ on a heterogeneous graph $G$:  \textbf{Relation Type Usefulness Distributions (RTUD)} is a group of task-preferred (usefulness) probability distributions over relation types, which is defined at graph schema level (global level) of $G$. As Figure \ref{fig:RTUD} shows, RTUD can be represented as a probability matrix $\beta$ of size $\left |\mathbb{N}  \right | \times \left |\mathbb{Z}  \right |$, where $i^{th}$ row of this matrix represents a relation type usefulness distribution $\beta_{i}$ given a specific vertex type $\mathbb{N}_i$, in which $\beta_{i,j} = Pr(\mathbb{Z}_{j}|\mathbb{N}_{i}) $ denotes the usefulness probability of a relation type $\mathbb{Z}_{j}$ given $\mathbb{N}_i$.

Correspondingly, \textbf{Relation Transition Distributions (RTD)} is a group of task-independent probability distributions associated to relations, which is defined at graph instance level (local level) of $G$. Given a vertex $v^{*}$ of type $\mathbb{N}_{n}$, $\alpha_{j}=Pr_{RTD}^{j}(V_{m}|v^{*})$ denotes the transition distribution of a type $\mathbb{Z}_{j}$ relation  (from vertex type $\mathbb{N}_{n}$ to vertex type $\mathbb{N}_{m}$), $V_{m}$ are vertexes of $\mathbb{N}_{m}$ type. RTD aims to reflect the basic semantics of different types of relations in $G$ and focuses on the local structure around $v^{*}$. For instance, Table \ref{tab:graph} defines the RTD of cross-language
heterogeneous graph.

As Figure \ref{fig:example} shows, with RTUD and RTD, we can simulate a hierarchical random walk of fixed length $l$ in $G$. In order to avoid walking into a dead end, the directions of relations are ignored in hierarchical random walk algorithm. The hierarchical random walk process is as follows:

\begin{enumerate}
\item For $i_{th}$ vertex $v_{i}$ in the walk:
    \begin{enumerate}
    \item Generate $\beta_{n} = (\beta_{n,1},\cdots ,\beta_{n,|\mathbb{Z}|})$ from RTUD based on $v_{i}$'s vertex type $\mathbb{N}_{n}$
    \item Probabilistically draw a relation type $\mathbb{Z}_{z}$ from $\beta_{n}$
    \item For the generated relation type $\mathbb{Z}_{z}$ 
        \begin{enumerate}
        \item Generate $\alpha_{z}$ from RTD based on $\mathbb{Z}_{z}$
        \item Based on $v_{i}$, probabilistically draw one vertex from $\alpha_{z}$ as the destination vertex $v_{i+1}$ 
        \item Walk forward to $v_{i+1}$ 
        \end{enumerate}
    \end{enumerate}
\end{enumerate}

\begin{algorithm}
\small
\caption{RTUD training algorithm: a K-shortest paths ranking based EM approach}  
\label{alg:em}  
\begin{algorithmic}[1]  
\State {Initialize RTUD, a.k.a, the probability matrix $\beta$ (Row: vertex type, column: relation type). For each row $\beta_{i}$, every element $\beta_{i,j}$ is set to be equal} (here, $\beta_{i,j}$ denotes $Pr(\mathbb{Z}_{j}|\mathbb{N}_{i})$, must fit for the graph schema $S_{G}=(\mathbb{N},\mathbb{Z})$, the $\beta_{i,j}$ that violates $S_{G}$ is set to 0)
\\\hrulefill
\Procedure{E-Step: K-shortest paths ranking}{}
    \State{Initialize $\Theta$, every element is set to be zero}
    \For{each $\left \{ v_{s}, v_{t} \right \} \in Set_{L}$}
        \State{Calculate $w(p)$ and find K-shortest paths $P^{*}$}
        \For{each $p^{*} \in P^{*}$}
            \For{each relation $e \in \mathbb{Z}_{j}$ from $p^{*}$}
                \State{Update $\Theta_{j}$ by $\Theta_{j}= F_{\Theta}(c)$  }
            \EndFor
        \EndFor
    \EndFor
    \State{ Call \Call{M-Step}{}}
\EndProcedure
\\\hrulefill
\Procedure{M-Step: Update $\beta$ }{}
    \For{each row $\beta_{i} \in \beta$ (vertex type)}
        \For{each relation $\mathbb{Z}_{j} \in \mathbb{Z} $}
            \State{update $\beta_{i,j}$ by $\beta_{i,j}^{n+1} = F_{\beta}(\beta_{i,j}^{n},\Theta)$}
        \EndFor
    \EndFor
    \If{$P_{All}^{*}$ stabilize (a.k.a, $\varepsilon$\% of the shortest paths ranking in the all shortest path sets are no longer changing)} 
        \State{Algorithm End} 
 	\Else 
    	\State{ Call \Call{E-Step}{}}
 	\EndIf
\EndProcedure
\end{algorithmic}
\end{algorithm}

In this study, with a set of $M$ labeled vertex pairs $Set_{L}$, we propose an iterative K-shortest paths ranking based EM (expectation - maximization) approach to obtain and optimize RTUD. $Set_{L}$ is generated based on the task-specified relevance. For instance, for CCR task, a pair of paper vertexes $\left \{ v_{s}, v_{t} \right \}$ connected via a cross-language relation could be a labeled pair for RTUD training. For a specific task, the representations of relevant pair of vertexes should be similar. 

Then, in the proposed hierarchical representation learning framework, the goals are: (1) the vertex neighborhood $N_{v}$ should contain the task-relevant vertexes to the greatest extent possible; (2) the distance (random walk sequence length) between two task-relevant vertexes should be as short as possible. In other words, RTUD should be trained to navigate the random walk between the related vertexes pairs, a.k.a., with the trained RTUD, there is a greater chance that one relevant vertex could random walk to another relevant one on the heterogeneous graph.

We formalize this goal as a K-shortest paths ranking problem. Let a path $p$ from $v_{s}$ to $v_{t}$ in $G$ is a sequence of vertexes and relations with the form: 
$$p=\left \{ v_{s}\overset{relation_{1}}{\rightarrow}  v_{s+1} \cdots  \overset{relation_{i}}{\rightarrow} v_{t} \right \}$$ 
$P$ denotes the set of all paths from $v_{s}$ to $v_{t}$ in $G$. Given a relation $e_{r}$ of type $\mathbb{Z}_{z}$ from vertex $v_{i}$ of type $\mathbb{N}_{n}$ to vertex $v_{j}$ of type $\mathbb{N}_{m}$, the $e_{r}$'s weight $w_{i,j}^{z}$ integrated RTUD and RTD, which can be calculated as:
$$w_{i,j}^{z} = \frac{1}{Pr_{RTUD}(\mathbb{Z}_{z}|\mathbb{N}_{n})\cdot Pr_{RTD}^{z}(v_{j}|v_{i})}$$
The weight function of $p$ is $w(p) = \sum _{p} w_{i,j}^{z}$, a weight sum of all relations from $p$. The shortest path objective is the determination of a path $p^{*} \in P$ for which $w(p^{*}) \leq w(p)$ holds for any path $p \in P$ \cite{gallo1986shortest}. Then, the K-shortest paths objective is extended to determine the second, third,..., Kth shortest paths in $P$, that can be denoted as $P^{*}$. There are lots of efficient algorithms for this problem, we utilize the method proposed in \cite{de1990shortest}. The RTUD training utilizes an EM framework, as described in Algorithm \ref{alg:em}.

In Algorithm \ref{alg:em}, $\Theta = \left \{ \Theta_{1},\cdots ,\Theta_{|\mathbb{Z}|}) \right \} $ is a relation type update factor vector, $\Theta_{i}$ denotes the update value of $i$th relation type, $c$ denotes one count for the appearance of a specific type relation in the shortest paths. We explore 3 ways for relation type update factor function $F_{\Theta}$:

\textbf{Raw Count (RC)}: $F_{\Theta} = c++$. During each iteration, directly accumulate the relation type count.

\textbf{Length-Normalized Count (LNC)}: $F_{\Theta} = \frac{c}{L_{p^{*}}} ++$. During each iteration, accumulate the relation type count that is normalized by the path length.  $L_{p^{*}}$ is the length of path $p^{*}$. By doing so, we try to minimize the possible bias from the long paths.

\textbf{Log-Discounted Count (LDC)}: $F_{\Theta} = \frac{c}{log_{2}(k+1)} ++$. During each iteration, accumulate the relation type count that is discounted by path ranking.  $K$ is the rank of the path $p$, the shortest path's rank is 1. Using this update function, different shortest paths are given different weights.

For RTUD update function $F_{\beta}$, we define 2 different forms:

\textbf{Direct Sum (DS)}: in DS, we update $\beta$ by directly adding the update values, $\eta$ is for normalization, $F_{\beta} = \frac{\beta_{i,j} +\Theta_{j}}{\eta }$.

\textbf{Sum with a Dumping Factor (SDF)}: in order to avoid the extreme probability, in SDF we add a dumping factor $\lambda$ for updating, $|\mathbb{Z}|^{*}$ is the possible relation type amount for a specific vertex type (constrained by graph schema). 
$$F_{\beta} = ( \lambda  (\frac{\beta_{i,j} +\Theta_{j} }{\sum _{|\mathbb{Z}|} (\beta_{i,j} +\Theta_{j}) })   + \frac{1-\lambda}{|\mathbb{Z}|^{*}})/ \eta $$

Note that, RTUD is constrained by graph schema. Given a vertex type $\mathbb{N}_{i}$, if a relation type $\mathbb{Z}_{j}$ violates the graph schema, the probability $Pr(\mathbb{Z}_{j}|\mathbb{N}_{i})$ will be set to zero. For instance, for a keyword vertex, the usefulness probability of the paper citation relation (a relation between paper vertex pairs) is zero. RTUD is task-specified, that means RTUD can dynamically change for different tasks, even though they share the same graph (e.g., we can use this graph for collaborator recommendation task, but the corresponding RTUD may change).

The pseudocode for Hierarchical Representation Learning on Heterogeneous Graph (HRLHG) is given in Algorithm \ref{alo:HRLHG}. By applying $r$ random walks of fixed length $l$ starting from each vertex in $G$, we can minimize the implicit random walk biases. The RTUD $\beta$ can be pre-trained by the K-shortest paths ranking based EM approach. The space complexity of HRLHG is $O(|E|)$, where $|E|$ is the relation number of $G$. The time complexity is $O(|\mathbb{Z}|+D)$ per hierarchical random walk, where  $|\mathbb{Z}|$ is the relation type number, and $D$ is the relation instance number of a specific sampled type connected to the current walking vertex. The time complexity can be further reduced, as suggested by \cite{grover2016node2vec}, if we parallelize the hierarchical random walk simulations, and execute them asynchronously\footnote{The source code of HRLHG, constructed graph data (with labeled ground truth) and learned representations are available at https://github.com/GraphEmbedding/HRLHG\label{foot:web}}.

\begin{algorithm}[htbp]
\small
\caption{Hierarchical Representation Learning on Heterogeneous Graph (HRLHG)}
\label{alo:HRLHG}
\begin{algorithmic}[1]
\State \textbf{RepresentationLearning} (Heterogeneous Graph $G=(V,E,\tau, \gamma )$, Relation Transition Distributions (RTD) $\alpha$, Dimensions $d$, Walks per vertex $r$,  Random Walk Length $l$, Context Window size $ws$, Task-specified Relevance Labeled Set $Set_{L}$)
\State $\beta$ = \textbf{K-ShortestPathEM} ($G$,$Set_{L}$)
\State Initialize \textit{walks} to Empty
\For{\textit{iter = 1} to $r$}
    \ForAll{vertexes $v \in V$}
        \State \textit{walk} = \textbf{HierarchicalRandomWalk} ($G,v,l,\beta,\alpha$)
        \State \textbf{Append} \textit{walk} to \textit{walks}
    \EndFor
\EndFor
\State $f=$ \textbf{HeterogeneousSkipGram} $(ws,d,walks)$
\State \Return \textit{f}
\\\hrulefill
\State \textbf{HierarchicalRandomWalk} (Heterogeneous Graph $G=(V,E,\tau, \gamma )$, Start vertex $v$, Random Walk Length $l$, Relation Type Usefulness Distributions (RTUD) $\beta$, Relation Transition Distributions (RTD)  $\alpha$)
\State Initialize $walk$ to $\left \{ v \right \}$
\For{$walk\_step=1$ to $l$}
    \State{Generate $\beta_{n}$ from $\beta$ based on $v$'s vertex type $\mathbb{N}_{n}$}
    \State{Probabilistically draw a relation type $\mathbb{Z}_{z}$ from $\beta_{n}$}
    \State{Based on $v$, probabilistically draw one vertex  $v_{t}$ from $\alpha_{z}$}
    \State{Append $v_{t}$ to $walk$}
\EndFor
\State \Return $walk$
\end{algorithmic}
\end{algorithm}

%% file: experiment.tex
\vspace{-2ex}\section{Experiment}

\subsection{Dataset and Experiment Setting}\label{ssec:data}
\textbf{Dataset}\textsuperscript{\ref{foot:web}}. We validated the proposed approach in a citation recommendation task between Chinese and English digital libraries. The goal was to recommend English candidate cited papers for a given Chinese publication. For this experiment, we collected 14,631 Chinese papers from the Wanfang digital library and 248,893 English papers from the Association for Computing Machinery (ACM) digital library (both in computer science). There were 750,557 English-to-English paper citation relations, 11,252 Chinese-to-Chinese paper citation relations, 27,101 Chinese-to-English paper citation relations, and 12,403 English papers had been cited by 7,900 Chinese papers. By using machine translation\footnote{As this study is not focusing on machine translation, we use Google translation API (https://cloud.google.com/translate) to translate the paper abstract and keywords.\label{foot:mt}} and language modeling (with Dirichlet smoothing). We generated 158,000 cross-language semantic matching relations (from Chinese to English). There were 3,953 Chinese keywords associated to the collected Chinese papers, and the Chinese paper-keyword-associated relation number was 7,316; while there were 7,436 English keywords associated to the collected English papers and the English paper-keyword-associated relation number was 903,265. Between keywords, There were 283,268 English-to-English keyword citation relations, 2,973 Chinese-to-Chinese keyword citation relations, 9,828 Chinese-to-English keyword citation relations. 2,564 Chinese keywords could be successfully translated into the corresponding English keywords\textsuperscript{\ref{foot:mt}}.

\textbf{Ground Truth and Evaluation Metric}. For evaluation, we generated a number of positive and negative instances to compare different algorithms for CCR task. The actual cross-language citation relation was used as ground truth (as 0 or 1 relevant scores) for evaluation. For example, if a candidate ACM paper was cited by the a testing Wanfang paper, the relevant score was 1, otherwise it was 0. We generated test and candidate collection data by using the following method: (1) randomly selected a certain proportion of papers from 7,900 Chinese papers that had cross-language citation relations to English corpus; (2) removed all cross-language citation relations from selected Chinese papers (Other relations, e.g., Chinese citation relations, were kept for model training); (3) the selected papers were used as a test collection. All the English papers cited by the Chinese papers in the test collection were used as candidate (cited paper) collection. For evaluation, the different models were compared by using the mean average precision (MAP), normalized discounted cumulative gain at rank (NDCG), precision (P) and Mean Reciprocal Rank (MRR).

\textbf{Validation Set}. For HRLGH, there were several hyper parameters (i.e., $k$ for shortest path EM method) and algorithm functions (i.e., $F_{\Theta}$ and $F_{\beta}$ for RTUD training) needed to be tuned. Meanwhile, for a fair comparison, we also tuned the hyper parameters of baselines (i.e., return parameter $p$ and in-out parameter $q$ for node2vec algorithm) for making sure the baseline algorithms could achieve the best performance. So, we constructed a validation set following the process described above (10\% papers were randomly selected for validation). A comprehensive model component analysis and baseline hyper parameter tuning would be conducted via this validation set.

\textbf{Baselines}. We compared with three groups of representation algorithms, from text or graph viewpoints, to comprehensively evaluate the performance of the proposed method. 10-fold cross-validation was applied to avoid evaluation bias. 

\textit{Textual Content Based Method}.

1. Embedding Transformation~\cite{mikolov2013exploiting}: We transformed the testing Chinese paper's abstract embedding into the English embedding space through a trained transformation matrix. Then, recommend the  English citations based on the transformed Chinese abstract embedding, denoted as \textbf{EF}.

2. Machine Translation by Google Translation API + Language Model (with Dirichlet smoothing)~\cite{zhai2001study}: We translated the testing Chinese paper's abstract into English, and then used language model to recommend English citations, denoted as \textbf{MT+LM}.

\textit{Collaborative Filtering Based Method}.

3. Item-based Collaborative Filtering~\cite{sarwar2001item}: Recommended English citations using Item-based Collaborative Filtering based on (monolingual + cross-language) citation relations, denoted as \textbf{CF$_{I}$}.

4. Popularity-based Collaborative Filtering~\cite{su2009survey}: Recommended English citations using Popularity-based Collaborative Filtering based on (monolingual + cross-language) citation relations, denoted as \textbf{CF$_{P}$}.

\textit{Network Representation Learning Based Method}.

5. DeepWalk~\cite{perozzi2014deepwalk}: We used DeepWalk to learn the graph embeddings via \textit{uniform random walk} in the network and recommended English citations based on the learned embeddings. Because DeepWalk was originally designed for homogeneous graph, for a fair comparison, we applied DeepWalk on two graphs, (1) citation network, denoted as \textbf{DW$_{c}$}; (2) all typed networks with accumulated relation weights. For this approach, we integrated all relations between two vertexes into one edge, and the weight was estimated by the sum of all integrated relations. Then, a heterogeneous graph could be simplified to a homogeneous graph, denoted as \textbf{DW$_{all}$}.

6. LINE~\cite{tang2015line}: LINE aimed at preserving first-order and second-order proximity in concatenated embeddings. Similar as DeepWalk, we applied LINE on two graphs, with LINE 1st-order and 2nd-order representation approaches. So, there were four different baseline models, denoted as \textbf{LINE$_{c}^{1st}$}, \textbf{LINE$_{c}^{2nd}$}, \textbf{LINE$_{all}^{1st}$} and \textbf{LINE$_{all}^{2nd}$}.

7. node2vec~\cite{grover2016node2vec}: node2vec learned graph embeddings via \textit{2nd order random walks} in the network. Similar as DeepWalk and LINE, we also applied node2vec on two graphs for comparison, denoted as \textbf{N2V$_{c}$} and \textbf{N2V$_{all}$}. We tuned return parameter $p$ and in-out parameter $q$ with a grid search over $p, q \in \left \{ 0.25, 0.50, 1, 2, 4 \right \}$ on the validation set, and picked up a best performed parameter setting for experiment, as suggested by~\cite{grover2016node2vec}.

8. metapath2vec++ \cite{dong2017metapath2vec}: metapath2vec++ was originally designed for heterogeneous graphs. It learned heterogeneous graph embeddings via \textit{metapath based random walk} and \textit{heterogeneous negative sampling} in the network. Metapath2vec++ required a human-defined metapath scheme to guide random walks. We tried 3 different metapaths for this experiment: (1) $P_{s} \overset{h}{\rightarrow} K_{s} \overset{c}{\rightarrow} K_{t} \overset{h}{\leftarrow} P_{t}$, (2) $P_{s} \overset{h}{\rightarrow} K_{s} \overset{t}{\rightarrow} K_{t} \overset{h}{\leftarrow} P_{t}$, (3) $P_{s} \overset{s}{\rightarrow} P_{t} \overset{c}{\rightarrow} P_{t}$. These metapaths were denoted as \textbf{M2V++$_{1}$}, \textbf{M2V++$_{2}$} and \textbf{M2V++$_{3}$}, respectively. We also trained two learning to rank models (Coordinate Ascent \cite{metzler2007linear} and ListNet \cite{cao2007learning}) to further integrate these three metapath2vec++ models (by utilizing each metapath as a ranking feature), denoted as \textbf{M2V++$_{CA}$} and \textbf{M2V++$_{LN}$}.

For a fair comparison, for all the random walk based embedding methods, we used the same parameters as follows: (1)  The number of walks per vertex $r$: 10; (2) the walk length $l$: 80; (3) the vector dimension $d$: 128; (4)  the neighborhood size (Context Window size) $ws$: 10. Please note that most original baseline papers used the above parameter settings, and the proposed method also shared the same parameters. For the experiment fairness, we didn't tune those parameters on validation set. We applied the parameter sensitivity analysis in section \ref{ssec:para}.



\vspace{-2ex}\subsection{Impact of Different Model Components}\label{ssec:para}

\begin{figure}[htbp]\centering
 	\includegraphics[width=1.0\columnwidth]{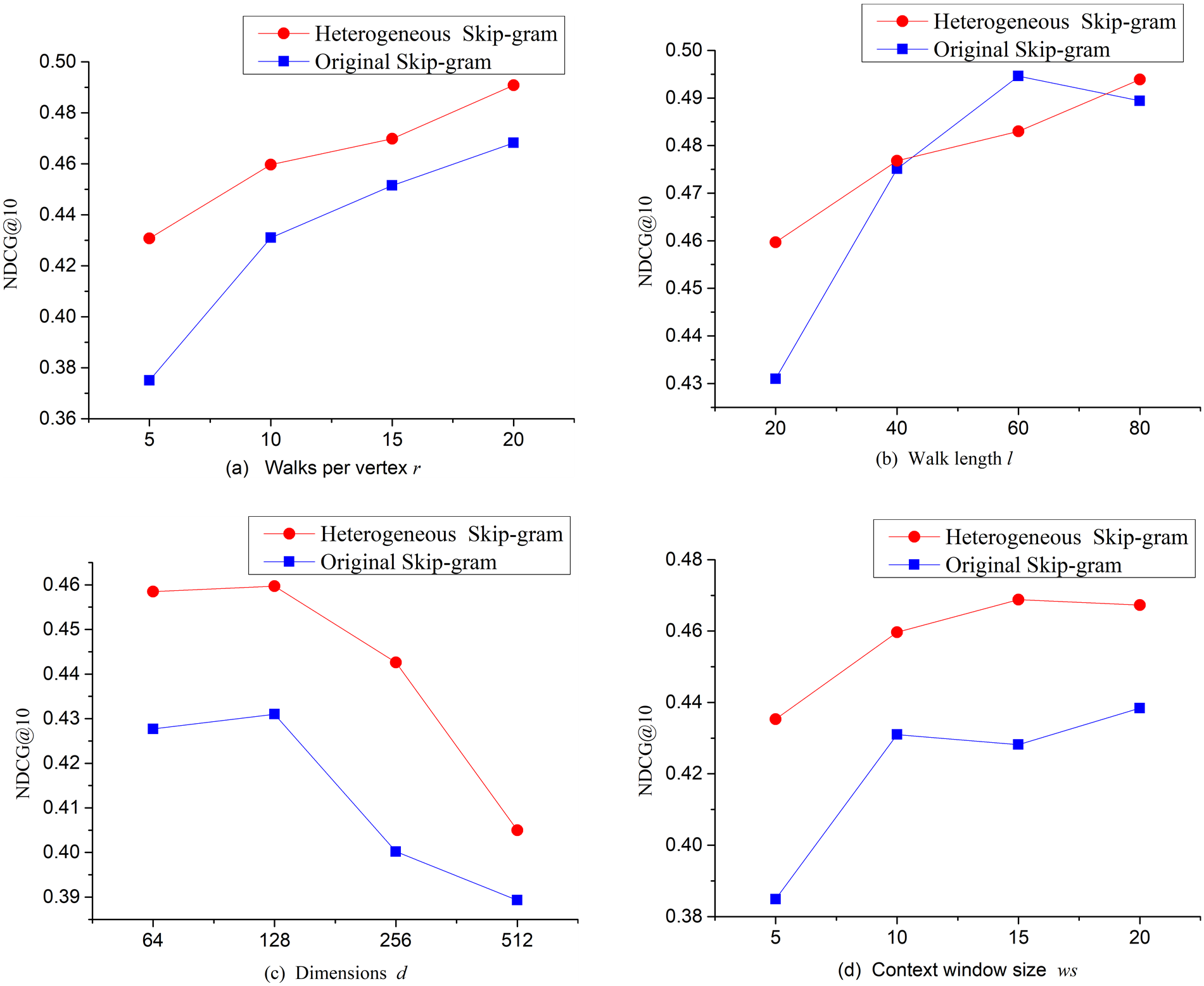}\vspace{-2ex}
 	\caption{Sensitivity analysis for the embedding related parameters}
 	\label{fig:para}
\end{figure}

\begin{figure*}[htbp]\centering
 	\includegraphics[width=1.8\columnwidth]{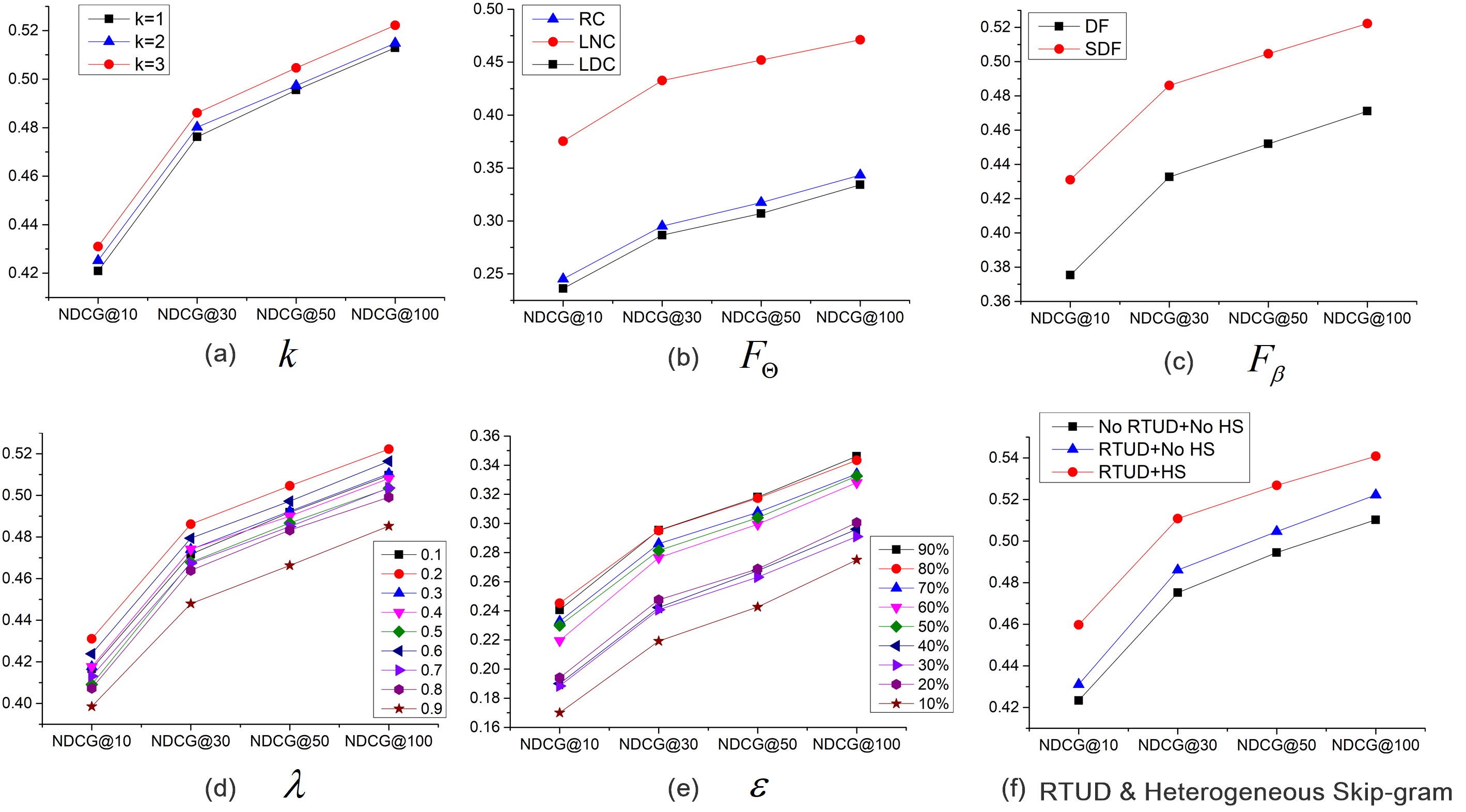}\vspace{-2ex}
 	\caption{Hyper parameter comparison and algorithm component validation for the proposed method: (a) Comparison of $k$ for K-shortest paths ranking based EM training; (b) Comparison of relation type update factor function $F_{\Theta}$ for RTUD training; (c) Comparison of RTUD update function $F_{\beta}$ for training; (d) Comparison of the dumping factor $\lambda$ for $F_{\beta}$ of SDF form; (e) Comparison of the convergence percentage $\varepsilon$ of K-shortest paths ranking based EM algorithm; (f) Validation of relation type usefulness distributions (RTUD) and heterogeneous skip-gram (HS). (The embedding related parameter setting is: the number of walks per vertex $r=10$; the walk length $l=20$; the vector dimension $d=128$; the context window size $ws=10$)}
 	\label{fig:hyper}
\end{figure*}

\begin{table*}[htbp]
\centering
\small
\caption{Measures of different cross-language citation recommendation algorithms}\vspace{-2ex} 
\label{tab:result}
\begin{threeparttable}
\begin{tabular}{l l l l l l l l l l l}
\hline
\textbf{Algorithm} & \textbf{NDCG@10} & \textbf{NDCG@30} & \textbf{NDCG@50} &\textbf{P@10} & \textbf{P@30} & \textbf{P@50} &\textbf{MAP@10} & \textbf{MAP@30}& \textbf{MAP@50} & \textbf{MRR} \\ \hline
EF        & 0.0176 & 0.0301 & 0.0384 & 0.0072 & 0.0060 & 0.0054 & 0.0101 & 0.0129 & 0.0140 & 0.0300  \\
MT+LM     & 0.3404 & 0.3811 & 0.3966 & 0.1225 & 0.0573 & 0.0387 & 0.2563 & 0.2739 & 0.2777 & 0.4343   \\ 
CF$_{I}$  & 0.0980 & 0.1034 & 0.1059 & 0.0330 & 0.0134 & 0.0086 & 0.0772 & 0.0793 & 0.0796 & 0.1290  \\ 
CF$_{P}$  & 0.0041 & 0.0082 & 0.0108 & 0.0026 & 0.0024 & 0.0022 & 0.0017 & 0.0023 & 0.0026 & 0.0090  \\ 
DW$_{c}$  & 0.2713 & 0.3060 & 0.3177 & 0.1037  & 0.0502  & 0.0336  & 0.2162  & 0.2348  & 0.2381  &  0.3053  \\ 
DW$_{all}$ & 0.3606  & 0.4214  & 0.4416  & 0.1463  & 0.0735  & 0.0499 & 0.2679  & 0.2979  & 0.3033  & 0.4077  \\ 
LINE$_{c}^{1st}$  & 0.2258  & 0.2557  & 0.2674  & 0.0854  & 0.0421  & 0.0289  & 0.1777  & 0.1927  & 0.1958  & 0.2628   \\ 
LINE$_{c}^{2nd}$  & 0.1499  & 0.1730  & 0.1822  & 0.0572  & 0.0295  & 0.0205  & 0.1136  & 0.1241  & 0.1263  & 0.1894   \\
LINE$_{all}^{1st}$  & 0.3534  & 0.4096   & 0.4302  & 0.1386  & 0.0691  & 0.0473  & 0.2671  & 0.2936  & 0.2990  & 0.4090   \\ 
LINE$_{all}^{2nd}$  & 0.1047  & 0.1385  & 0.1564  & 0.0453   & 0.0284  & 0.0221  & 0.0663  & 0.0775  & 0.0811  & 0.1544    \\
N2V$_{c}$  & 0.2724  & 0.3040  & 0.3153  & 0.1025  & 0.0489  & 0.0327  & 0.2183  & 0.2353  & 0.2383  & 0.3083   \\ 
N2V$_{all}$  & 0.4651  & 0.5194  & 0.5354  & 0.1730  & 0.0809  & 0.0533  & 0.3661   & 0.3951  & 0.3999  & 0.5194   \\
M2V++$_{1}$  & 0.0195   & 0.0214   & 0.0225   & 0.0052  & 0.0023  & 0.0015  & 0.0144  & 0.0147  & 0.0148  & 0.0312   \\ 
M2V++$_{2}$  & 0.0015  & 0.0031  & 0.0045  & 0.0006  & 0.0006  & 0.0006  & 0.0006  & 0.0007  & 0.0009  & 0.0037   \\ 
M2V++$_{3}$  & 0.0687  & 0.0933  & 0.1058  & 0.0308  & 0.0195  & 0.0150  & 0.0409  & 0.0481  & 0.0503  & 0.1070   \\ 
M2V++$_{LN}$  & 0.0198  & 0.0273  & 0.0321  & 0.0084  & 0.0054  & 0.0045   & 0.0113  & 0.0130  & 0.0135  & 0.0389   \\
M2V++$_{CA}$  & 0.0243  & 0.0335   & 0.0380   & 0.0107   & 0.0068   & 0.0052    & 0.0136   & 0.0156   & 0.0161   & 0.0451    \\
\textbf{HRLHG} & \textit{\textbf{0.5034}}\textsuperscript{\dag\dag\dag}  & \textit{\textbf{0.5522}}\textsuperscript{\dag\dag\dag} & \textit{\textbf{0.5664}}\textsuperscript{\dag\dag\dag} & \textit{\textbf{0.1840}}\textsuperscript{\dag\dag\dag} & \textit{\textbf{0.0832}}\textsuperscript{\dag\dag\dag} & \textit{\textbf{0.0543}}\textsuperscript{\dag\dag\dag} & \textit{\textbf{0.4033}}\textsuperscript{\dag\dag\dag} & \textit{\textbf{0.4309}}\textsuperscript{\dag\dag\dag} & \textit{\textbf{0.4353}}\textsuperscript{\dag\dag\dag} & 
\textit{\textbf{0.5598}}\textsuperscript{\dag\dag\dag}\\ \hline 
\end{tabular}  
\begin{tablenotes}
    \footnotesize
    \item Significant test: \textsuperscript{\dag}$p < 0.01$, \textsuperscript{\dag\dag}$p < 0.001$, \textsuperscript{\dag\dag\dag}$p < 0.0001$
    \end{tablenotes}
\end{threeparttable}
\end{table*}

For the proposed HRLHG, there were several important parameters and functions. To explore the effects of those model components, on the validation set, we compared the cross-language recommendation performances of proposed method under different model settings (by varying the examined model component while kept others fixed). We mainly focused on following components: (a) parameter $k$ for K-shorest paths based EM algorithm, we compared and selected the best $k$ from $k \in \left \{ 1,2,3 \right \}$. (b) $F_{\Theta}$, relation type update factor function for RTUD training: Raw Count (RC), Length-Normalized Count (LNC) and Log-Discounted Count (LDC). (c) $F_{\beta}$, RTUD update function for model training: Direct Sum (DS) and Sum with a Dumping Factor (SDF). (d) $\lambda$, parameter for $F_{\beta}$ of SDF form, we compared and selected the best $\lambda$ from $\lambda \in \left \{ 0.1, 0.2,...,0.9 \right \}$. (e) $\varepsilon$, the convergence percentage of EM algorithm, we tried $\varepsilon$ over 90\% to 10\%. (f) Validation of relation type usefulness distributions (RTUD) and heterogeneous skip-gram (HS). 

Note that, we conducted a comprehensive comparison experiment. For each examined model component, we tried multiple combinations of other components to avoid the possible bias brought by the component setting choices. For instance, we tested the impact of different $k$ under component combination ($F_{\Theta}$ = LNC, $F_{\beta}$= DS and $\varepsilon$= 80\% for RTUD training, while using ordinary skip-gram for embedding) and component combination ($F_{\Theta}$ = RC, $F_{\beta}$= SDF with $\lambda$ = 0.8 and $\varepsilon$= 20\% for RTUD training, while using heterogeneous skip-gram for embedding), respectively. Because of the space limitation, we cannot report all results in this paper. The representative results on the validation set in terms of NDCG are depicted in Figure \ref{fig:hyper} (the other comparison groups showed the similar trends).

As we can see, considering more shortest paths in RTUD training brings a performance improvement. Length-normalized count (LNC) function could achieve best among the three $F_{\Theta}$ choices. For $F_{\beta}$, sum with a dumping factor (SDF) outperforms direct sum (DS). If we utilized SDF, a small $\lambda$ could be superior than a great one. A possible explanation was that a small $\lambda$ would penalize the dominated relation type usefulness probability to avoid overfitting. Generally, the algorithm performed better when more $P^{*}$ became stabilize in training iterations.

To validate the effectiveness of RTUD and heterogeneous skip-gram (HS), we also compared the performance of our model without them. As Figure \ref{fig:hyper} (f) showed, when RTUD was removed (treating each relation type equally when we conducted the hierarchical random walk) and HS was replaced by ordinary skip-gram, recommendation performance declined significantly. It is clear that RTUD and HS contribute to heterogeneous graph based random walk and recommendation performance significantly. More importantly, RTUD doesn't need any human intervention or expert knowledge. 

Based on the comparison and analysis, we selected a component setting ($k$=3, $F_{\Theta}$ = LNC, $F_{\beta}$= SDF with $\lambda$ = 0.2 and  $\varepsilon$= 80\% for RTUD training, while using heterogeneous skip-gram for vertex embedding), for further experiments with Baselines.

In skip-gram-based representation learning models, there were several common parameters (see Section \ref{ssec:data}). We also conducted a sensitivity analysis of HRLHG to these parameters. Figure \ref{fig:para} showed the their impacts on recommendation performance.

\vspace{-2ex}\subsection{Comparison with Baselines}

The cross-language citation recommendation performance results of different models were displayed in Table \ref{tab:result}. Based on the experiment results, we had the following observations: (1) The proposed method significantly outperformed ($p<0.0001$) other baseline models for all evaluation metrics. For instance, in terms of MAP@10, HRLHG achieved at least 10\% improvement, comparing with all other 17 baselines. (2) The traditional models solely relied on one kind of information, i.e., machine translation based methods (EF, MT+LM) or citation relation based collaborative filtering approaches (CF$_{I}$ and CF$_{P}$) cannot work as well as other network embedding based methods. (3) Although designed for homogeneous networks, by adding more types of vertexes and relations, the performance of DeepWalk (DW$_{all}$), LINE (LINE$_{all}^{1st}$ and LINE$_{all}^{2nd}$) and node2vec (N2V$_{all}$) had significant improvements over the ones using only citation networks (DW$_{c}$, LINE$_{c}^{1st}$, LINE$_{c}^{2nd}$ and N2V$_{c}$). This observation confirmed that heterogeneous information did enhance the models' representation learning abilities. (4) metapath2vec++, designed for heterogeneous information networks, didn't work well in the experiment. Even after applying learning to rank algorithms for integrating multiple metapath2vec++ models, the recommendation results were still not good. A possible explanation was that, for CCR task, no single metapath could cover the recommendation requirement. In addition, metapath based random walk was too strict to explore potential useful neighbourhoods for vertex representation learning. This observation also indicated that metapath2vec++ was depending on domain expert knowledge. If one cannot find the optimize metapath, the embedding performances were even worse than the homogeneous network representation learning models. 

For each vertex type, HRLHG trained a relation type usefulness distribution. Based on the learned RTUD (available at project website), we can obtain the task-specified knowledge and improve the interpretability of proposed graph representation model. For instance, in the experimental CCR task, when a random walker reaches an English vertex, for the next move, the probabilities of $P_{E} \overset{c}{\rightarrow} P_{E}$ (an English paper cites another English paper) and $P_{E} \overset{h}{\rightarrow} K_{E}$ (an English paper has an English keyword) are higher than other relation types. This distribution navigates the walker to prefer to stay in the English repository rather than going back to the Chinese repository. By conducting the hierarchical random walk based on RTUD, the task specific knowledge can be further embedded into the representations learned by HRLHG.

In sum, for the CCR task, the proposed HRLHG method could automatically learn the relation type usefulness distributions for random walk navigation and the new method significantly outperformed the current text, homogeneous graph and heterogeneous graph embedding methods.

%% file: review.tex
\vspace{-2ex}\section{RELATED WORK}\label{review}

Citation recommendation aims to recommend a list of citations (references) based on the similarity between the recommended papers and user profiles or samples of in-progress text. For instance, He et al.~\cite{he2010context} proposed a probabilistic model to compute the relevance score based on contexts of a citation and its abstract. Jiang et al.~\cite{jiang2015chronological} generated a heterogeneous graph with various relations between topics and papers, and a supervised random walk was used for citation recommendation. From bibliographic viewpoint, Shi, Leskovec, and McFarland~\cite{citing-impact} developed citation projection graphs by investigating citations among publications cited by a given paper. Collaborative filtering algorithm can also be used for recommending citation papers \cite{mcnee2002recommending}. However, all of the prior studies focused on monolingual citation recommendation and cannot be directly used for cross-language citation recommendation. Intuitively, translation-based models can be addressed for cross-language recommendation. Recently, word embedding is a powerful approach for content representation \cite{mikolov2013efficient}. Mikolov et al. \cite{mikolov2013exploiting} transformed one language's vector space into the space of another by utilizing a linear projection with a transformation matrix $W$. This approach is effective for word translation, but the translation effect for scholarly text has not yet been demonstrated. Tang et al.~\cite{tang2014cross} proposed bilingual embedding algorithms, which were efficient for cross-language context-aware citation recommendation task. However, they ignored the important citation relations in their work.

Network embedding algorithms, namely graph representation learning models, which aim to learn the low-dimensional feature representations of nodes in networks, are attracting increasing attention recently. Based on the techniques utilized in the model, we can briefly classify these algorithms into the following categories: the graph factorization based models, e.g., GraRep \cite{cao2015grarep}; the shallow neural network based models, e.g., LINE \cite{tang2015line}; the deep neural network based models, e.g., GCN \cite{kipf2016semi}; and the random walk based method, e.g., DeepWalk \cite{perozzi2014deepwalk}, node2vec \cite{,grover2016node2vec} and metapathvec++ \cite{dong2017metapath2vec}. Technically the random walk based models are also using a shallow neural network. The main difference between random walk based models are the random walk algorithms used for generating the vertex sequences from the graph. A potential problem for GraRep and GCN is the space complexity ($O(N^2)$), and the computational costs of these models can be too expensive to embed the large complex networks in the real world. For instance, in this CCR experiment (a 200,000 vertexes level graph), the memory requirement of GraRep/GCN is over 600G.

In this study, we address the CCR problems and propose a novel method HRLHG to learn a mapping of publication to a low-dimensional joint embedding space for heterogeneous graph. HRLHG belongs to  the random walk based network embedding models. A hierarchical random walk is proposed to cope the task-specified problem on heterogeneous graph. To the best of our knowledge, few existing studies have investigated the graph embedding approach for cross-language citation recommendation problem. 

%% file: conclusion.tex
\section{Conclusion}

In this paper, we propose a new problem: cross-language citation recommendation (CCR). Unlike existing scholarly recommendation problem, CCR enables cross language and cross repository recommendation. The proposed Hierarchical Representation Learning on Heterogeneous Graph (HRLHG) model can project a publication into a joint embedding space, which encapsulate both semantic and topological information. By training a set of relation type usefulness distributions (RTUD) on a heterogeneous graph, we propose a hierarchical two-level random walk: the global level is for graph schema navigation (task-specific); while the local level is for graph instance (task-independent) walking. 

Unlike most prior heterogeneous graph mining methods, which employed expert-generated or rule-based ranking hypotheses to address recommendation problems, in a complex CCR graph, it can be difficult to exhaustively examine all of the potentially useful path types to generate metapaths. Furthermore, if a large number of random walk-based ranking functions are used, the computational cost can be prohibitive. Extensive experiments prove our hypothesis that the latent heterogeneous graph feature representations learned by HRLHG are able to improve cross-language citation recommendation performance (when comparing with 17 state-of-the-art baselines). In addition, the learned RTUD is able to reveal the latent task-specified knowledge, which is important to the interpretability of the proposed representation model. 

In the future, we will validate the proposed method on other heterogeneous graph embedding  based tasks, e.g., music recommendation or movie recommendation. Meanwhile, we will investigate more sophisticated method to generate RTUD. For instance, add personalization component to the algorithm, and enable personalized heterogeneous graph navigation for random walk optimization. 

%% file: Cross.bbl

\begin{thebibliography}{30}


\ifx \showCODEN    \undefined \def \showCODEN     #1{\unskip}     \fi
\ifx \showDOI      \undefined \def \showDOI       #1{#1}\fi
\ifx \showISBNx    \undefined \def \showISBNx     #1{\unskip}     \fi
\ifx \showISBNxiii \undefined \def \showISBNxiii  #1{\unskip}     \fi
\ifx \showISSN     \undefined \def \showISSN      #1{\unskip}     \fi
\ifx \showLCCN     \undefined \def \showLCCN      #1{\unskip}     \fi
\ifx \shownote     \undefined \def \shownote      #1{#1}          \fi
\ifx \showarticletitle \undefined \def \showarticletitle #1{#1}   \fi
\ifx \showURL      \undefined \def \showURL       {\relax}        \fi
\providecommand\bibfield[2]{#2}
\providecommand\bibinfo[2]{#2}
\providecommand\natexlab[1]{#1}
\providecommand\showeprint[2][]{arXiv:#2}

\bibitem[\protect\citeauthoryear{Bahdanau, Cho, and Bengio}{Bahdanau
  et~al\mbox{.}}{2015}]%
        {bahdanau2014neural}
\bibfield{author}{\bibinfo{person}{Dzmitry Bahdanau},
  \bibinfo{person}{Kyunghyun Cho}, {and} \bibinfo{person}{Yoshua Bengio}.}
  \bibinfo{year}{2015}\natexlab{}.
\newblock \showarticletitle{Neural machine translation by jointly learning to
  align and translate}. In \bibinfo{booktitle}{\emph{Proceedings of the
  International Conference on Learning Representations (ICLR)}}.
\newblock


\bibitem[\protect\citeauthoryear{Bengio, Courville, and Vincent}{Bengio
  et~al\mbox{.}}{2013}]%
        {bengio2013representation}
\bibfield{author}{\bibinfo{person}{Yoshua Bengio}, \bibinfo{person}{Aaron
  Courville}, {and} \bibinfo{person}{Pascal Vincent}.}
  \bibinfo{year}{2013}\natexlab{}.
\newblock \showarticletitle{Representation learning: A review and new
  perspectives}.
\newblock \bibinfo{journal}{\emph{IEEE transactions on pattern analysis and
  machine intelligence}} \bibinfo{volume}{35}, \bibinfo{number}{8}
  (\bibinfo{year}{2013}), \bibinfo{pages}{1798--1828}.
\newblock


\bibitem[\protect\citeauthoryear{Cao, Lu, and Xu}{Cao et~al\mbox{.}}{2015}]%
        {cao2015grarep}
\bibfield{author}{\bibinfo{person}{Shaosheng Cao}, \bibinfo{person}{Wei Lu},
  {and} \bibinfo{person}{Qiongkai Xu}.} \bibinfo{year}{2015}\natexlab{}.
\newblock \showarticletitle{Grarep: Learning graph representations with global
  structural information}. In \bibinfo{booktitle}{\emph{Proceedings of the 24th
  ACM International on Conference on Information and Knowledge Management}}.
  ACM, \bibinfo{pages}{891--900}.
\newblock


\bibitem[\protect\citeauthoryear{Cao, Qin, Liu, Tsai, and Li}{Cao
  et~al\mbox{.}}{2007}]%
        {cao2007learning}
\bibfield{author}{\bibinfo{person}{Zhe Cao}, \bibinfo{person}{Tao Qin},
  \bibinfo{person}{Tie-Yan Liu}, \bibinfo{person}{Ming-Feng Tsai}, {and}
  \bibinfo{person}{Hang Li}.} \bibinfo{year}{2007}\natexlab{}.
\newblock \showarticletitle{Learning to rank: from pairwise approach to
  listwise approach}. In \bibinfo{booktitle}{\emph{Proceedings of the 24th
  international conference on Machine learning}}. ACM,
  \bibinfo{pages}{129--136}.
\newblock


\bibitem[\protect\citeauthoryear{de~Azevedo, Madeira, Martins, and
  Pires}{de~Azevedo et~al\mbox{.}}{1990}]%
        {de1990shortest}
\bibfield{author}{\bibinfo{person}{Jos{\'e}~Augusto de Azevedo},
  \bibinfo{person}{Joaquim Jo{\~a}o ER~Silvestre Madeira},
  \bibinfo{person}{Ernesto Q~Vieira Martins}, {and} \bibinfo{person}{Filipe
  Manuel~A Pires}.} \bibinfo{year}{1990}\natexlab{}.
\newblock \showarticletitle{A shortest paths ranking algorithm}. In
  \bibinfo{booktitle}{\emph{Proceedings of the Annual Conference of
  Associazione Italiana di Ricerca Operativa: Models and Methods for Decision
  Support (AIRO'90)}}. \bibinfo{pages}{1--8}.
\newblock


\bibitem[\protect\citeauthoryear{Dong, Chawla, and Swami}{Dong
  et~al\mbox{.}}{2017}]%
        {dong2017metapath2vec}
\bibfield{author}{\bibinfo{person}{Yuxiao Dong}, \bibinfo{person}{Nitesh~V
  Chawla}, {and} \bibinfo{person}{Ananthram Swami}.}
  \bibinfo{year}{2017}\natexlab{}.
\newblock \showarticletitle{metapath2vec: Scalable representation learning for
  heterogeneous networks}. In \bibinfo{booktitle}{\emph{Proceedings of the 23rd
  ACM SIGKDD International Conference on Knowledge Discovery and Data Mining}}.
  ACM, \bibinfo{pages}{135--144}.
\newblock


\bibitem[\protect\citeauthoryear{Fu, Lee, and Lei}{Fu et~al\mbox{.}}{2017}]%
        {fu2017hin2vec}
\bibfield{author}{\bibinfo{person}{Tao-yang Fu}, \bibinfo{person}{Wang-Chien
  Lee}, {and} \bibinfo{person}{Zhen Lei}.} \bibinfo{year}{2017}\natexlab{}.
\newblock \showarticletitle{HIN2Vec: Explore Meta-paths in Heterogeneous
  Information Networks for Representation Learning}. In
  \bibinfo{booktitle}{\emph{Proceedings of the 2017 ACM on Conference on
  Information and Knowledge Management}}. ACM, \bibinfo{pages}{1797--1806}.
\newblock


\bibitem[\protect\citeauthoryear{Gallo and Pallottino}{Gallo and
  Pallottino}{1986}]%
        {gallo1986shortest}
\bibfield{author}{\bibinfo{person}{Giorgio Gallo} {and}
  \bibinfo{person}{Stefano Pallottino}.} \bibinfo{year}{1986}\natexlab{}.
\newblock \showarticletitle{Shortest path methods: A unifying approach}.
\newblock \bibinfo{journal}{\emph{Netflow at Pisa}} (\bibinfo{year}{1986}),
  \bibinfo{pages}{38--64}.
\newblock


\bibitem[\protect\citeauthoryear{Grover and Leskovec}{Grover and
  Leskovec}{2016}]%
        {grover2016node2vec}
\bibfield{author}{\bibinfo{person}{Aditya Grover} {and} \bibinfo{person}{Jure
  Leskovec}.} \bibinfo{year}{2016}\natexlab{}.
\newblock \showarticletitle{node2vec: Scalable feature learning for networks}.
  In \bibinfo{booktitle}{\emph{Proceedings of the 22nd ACM SIGKDD International
  Conference on Knowledge Discovery and Data Mining}}. ACM,
  \bibinfo{pages}{855--864}.
\newblock


\bibitem[\protect\citeauthoryear{Guo, Fan, Ai, and Croft}{Guo
  et~al\mbox{.}}{2016}]%
        {guo2016deep}
\bibfield{author}{\bibinfo{person}{Jiafeng Guo}, \bibinfo{person}{Yixing Fan},
  \bibinfo{person}{Qingyao Ai}, {and} \bibinfo{person}{W~Bruce Croft}.}
  \bibinfo{year}{2016}\natexlab{}.
\newblock \showarticletitle{A deep relevance matching model for ad-hoc
  retrieval}. In \bibinfo{booktitle}{\emph{Proceedings of the 25th ACM
  International on Conference on Information and Knowledge Management}}. ACM,
  \bibinfo{pages}{55--64}.
\newblock


\bibitem[\protect\citeauthoryear{He, Pei, Kifer, Mitra, and Giles}{He
  et~al\mbox{.}}{2010}]%
        {he2010context}
\bibfield{author}{\bibinfo{person}{Qi He}, \bibinfo{person}{Jian Pei},
  \bibinfo{person}{Daniel Kifer}, \bibinfo{person}{Prasenjit Mitra}, {and}
  \bibinfo{person}{Lee Giles}.} \bibinfo{year}{2010}\natexlab{}.
\newblock \showarticletitle{Context-aware citation recommendation}. In
  \bibinfo{booktitle}{\emph{Proceedings of the 19th international conference on
  World wide web}}. ACM, \bibinfo{pages}{421--430}.
\newblock


\bibitem[\protect\citeauthoryear{Jiang, Liu, and Gao}{Jiang
  et~al\mbox{.}}{2015}]%
        {jiang2015chronological}
\bibfield{author}{\bibinfo{person}{Zhuoren Jiang}, \bibinfo{person}{Xiaozhong
  Liu}, {and} \bibinfo{person}{Liangcai Gao}.} \bibinfo{year}{2015}\natexlab{}.
\newblock \showarticletitle{Chronological Citation Recommendation with
  Information-Need Shifting}. In \bibinfo{booktitle}{\emph{Proceedings of the
  24th ACM International on Conference on Information and Knowledge
  Management}}. ACM, \bibinfo{pages}{1291--1300}.
\newblock


\bibitem[\protect\citeauthoryear{Kipf and Welling}{Kipf and Welling}{2016}]%
        {kipf2016semi}
\bibfield{author}{\bibinfo{person}{Thomas~N Kipf} {and} \bibinfo{person}{Max
  Welling}.} \bibinfo{year}{2016}\natexlab{}.
\newblock \showarticletitle{Semi-Supervised Classification with Graph
  Convolutional Networks}.
\newblock \bibinfo{journal}{\emph{arXiv preprint arXiv:1609.02907}}
  (\bibinfo{year}{2016}).
\newblock


\bibitem[\protect\citeauthoryear{Lao and Cohen}{Lao and Cohen}{2010}]%
        {lao}
\bibfield{author}{\bibinfo{person}{Ni Lao} {and} \bibinfo{person}{William~W
  Cohen}.} \bibinfo{year}{2010}\natexlab{}.
\newblock \showarticletitle{Relational retrieval using a combination of
  path-constrained random walks}.
\newblock \bibinfo{journal}{\emph{Machine learning}} \bibinfo{volume}{81},
  \bibinfo{number}{1} (\bibinfo{year}{2010}), \bibinfo{pages}{53--67}.
\newblock


\bibitem[\protect\citeauthoryear{Liu, Yu, Guo, and Sun}{Liu
  et~al\mbox{.}}{2014}]%
        {liu2014meta}
\bibfield{author}{\bibinfo{person}{Xiaozhong Liu}, \bibinfo{person}{Yingying
  Yu}, \bibinfo{person}{Chun Guo}, {and} \bibinfo{person}{Yizhou Sun}.}
  \bibinfo{year}{2014}\natexlab{}.
\newblock \showarticletitle{Meta-Path-Based Ranking with Pseudo Relevance
  Feedback on Heterogeneous Graph for Citation Recommendation}. In
  \bibinfo{booktitle}{\emph{Proceedings of the 23rd ACM International
  Conference on Conference on Information and Knowledge Management}}. ACM,
  \bibinfo{pages}{121--130}.
\newblock


\bibitem[\protect\citeauthoryear{McNee, Albert, Cosley, Gopalkrishnan, Lam,
  Rashid, Konstan, and Riedl}{McNee et~al\mbox{.}}{2002}]%
        {mcnee2002recommending}
\bibfield{author}{\bibinfo{person}{Sean~M McNee}, \bibinfo{person}{Istvan
  Albert}, \bibinfo{person}{Dan Cosley}, \bibinfo{person}{Prateep
  Gopalkrishnan}, \bibinfo{person}{Shyong~K Lam}, \bibinfo{person}{Al~Mamunur
  Rashid}, \bibinfo{person}{Joseph~A Konstan}, {and} \bibinfo{person}{John
  Riedl}.} \bibinfo{year}{2002}\natexlab{}.
\newblock \showarticletitle{On the recommending of citations for research
  papers}. In \bibinfo{booktitle}{\emph{Proceedings of the 2002 ACM conference
  on Computer supported cooperative work}}. ACM, \bibinfo{pages}{116--125}.
\newblock


\bibitem[\protect\citeauthoryear{Metzler and Croft}{Metzler and Croft}{2007}]%
        {metzler2007linear}
\bibfield{author}{\bibinfo{person}{Donald Metzler} {and}
  \bibinfo{person}{W~Bruce Croft}.} \bibinfo{year}{2007}\natexlab{}.
\newblock \showarticletitle{Linear feature-based models for information
  retrieval}.
\newblock \bibinfo{journal}{\emph{Information Retrieval}} \bibinfo{volume}{10},
  \bibinfo{number}{3} (\bibinfo{year}{2007}), \bibinfo{pages}{257--274}.
\newblock


\bibitem[\protect\citeauthoryear{Mikolov, Chen, Corrado, and Dean}{Mikolov
  et~al\mbox{.}}{2013a}]%
        {mikolov2013efficient}
\bibfield{author}{\bibinfo{person}{Tomas Mikolov}, \bibinfo{person}{Kai Chen},
  \bibinfo{person}{Greg Corrado}, {and} \bibinfo{person}{Jeffrey Dean}.}
  \bibinfo{year}{2013}\natexlab{a}.
\newblock \showarticletitle{Efficient estimation of word representations in
  vector space}.
\newblock \bibinfo{journal}{\emph{arXiv preprint arXiv:1301.3781}}
  (\bibinfo{year}{2013}).
\newblock


\bibitem[\protect\citeauthoryear{Mikolov, Le, and Sutskever}{Mikolov
  et~al\mbox{.}}{2013b}]%
        {mikolov2013exploiting}
\bibfield{author}{\bibinfo{person}{Tomas Mikolov}, \bibinfo{person}{Quoc~V Le},
  {and} \bibinfo{person}{Ilya Sutskever}.} \bibinfo{year}{2013}\natexlab{b}.
\newblock \showarticletitle{Exploiting similarities among languages for machine
  translation}.
\newblock \bibinfo{journal}{\emph{arXiv preprint arXiv:1309.4168}}
  (\bibinfo{year}{2013}).
\newblock


\bibitem[\protect\citeauthoryear{Mikolov, Sutskever, Chen, Corrado, and
  Dean}{Mikolov et~al\mbox{.}}{2013c}]%
        {mikolov2013distributed}
\bibfield{author}{\bibinfo{person}{Tomas Mikolov}, \bibinfo{person}{Ilya
  Sutskever}, \bibinfo{person}{Kai Chen}, \bibinfo{person}{Greg~S Corrado},
  {and} \bibinfo{person}{Jeff Dean}.} \bibinfo{year}{2013}\natexlab{c}.
\newblock \showarticletitle{Distributed representations of words and phrases
  and their compositionality}. In \bibinfo{booktitle}{\emph{Advances in neural
  information processing systems}}. \bibinfo{pages}{3111--3119}.
\newblock


\bibitem[\protect\citeauthoryear{Perozzi, Al-Rfou, and Skiena}{Perozzi
  et~al\mbox{.}}{2014}]%
        {perozzi2014deepwalk}
\bibfield{author}{\bibinfo{person}{Bryan Perozzi}, \bibinfo{person}{Rami
  Al-Rfou}, {and} \bibinfo{person}{Steven Skiena}.}
  \bibinfo{year}{2014}\natexlab{}.
\newblock \showarticletitle{Deepwalk: Online learning of social
  representations}. In \bibinfo{booktitle}{\emph{Proceedings of the 20th ACM
  SIGKDD international conference on Knowledge discovery and data mining}}.
  ACM, \bibinfo{pages}{701--710}.
\newblock


\bibitem[\protect\citeauthoryear{Ren, Liu, Yu, Khandelwal, Gu, Wang, and
  Han}{Ren et~al\mbox{.}}{2014}]%
        {ren2014cluscite}
\bibfield{author}{\bibinfo{person}{Xiang Ren}, \bibinfo{person}{Jialu Liu},
  \bibinfo{person}{Xiao Yu}, \bibinfo{person}{Urvashi Khandelwal},
  \bibinfo{person}{Quanquan Gu}, \bibinfo{person}{Lidan Wang}, {and}
  \bibinfo{person}{Jiawei Han}.} \bibinfo{year}{2014}\natexlab{}.
\newblock \showarticletitle{Cluscite: Effective citation recommendation by
  information network-based clustering}. In
  \bibinfo{booktitle}{\emph{Proceedings of the 20th ACM SIGKDD international
  conference on Knowledge discovery and data mining}}. ACM,
  \bibinfo{pages}{821--830}.
\newblock


\bibitem[\protect\citeauthoryear{Sarwar, Karypis, Konstan, and Riedl}{Sarwar
  et~al\mbox{.}}{2001}]%
        {sarwar2001item}
\bibfield{author}{\bibinfo{person}{Badrul Sarwar}, \bibinfo{person}{George
  Karypis}, \bibinfo{person}{Joseph Konstan}, {and} \bibinfo{person}{John
  Riedl}.} \bibinfo{year}{2001}\natexlab{}.
\newblock \showarticletitle{Item-based collaborative filtering recommendation
  algorithms}. In \bibinfo{booktitle}{\emph{Proceedings of the 10th
  international conference on World Wide Web}}. ACM, \bibinfo{pages}{285--295}.
\newblock


\bibitem[\protect\citeauthoryear{Shi, Leskovec, and McFarland}{Shi
  et~al\mbox{.}}{2010}]%
        {citing-impact}
\bibfield{author}{\bibinfo{person}{Xiaolin Shi}, \bibinfo{person}{Jure
  Leskovec}, {and} \bibinfo{person}{Daniel~A McFarland}.}
  \bibinfo{year}{2010}\natexlab{}.
\newblock \showarticletitle{Citing for high impact}. In
  \bibinfo{booktitle}{\emph{Proceedings of the 10th annual joint conference on
  Digital libraries}}. ACM, \bibinfo{pages}{49--58}.
\newblock


\bibitem[\protect\citeauthoryear{Su and Khoshgoftaar}{Su and
  Khoshgoftaar}{2009}]%
        {su2009survey}
\bibfield{author}{\bibinfo{person}{Xiaoyuan Su} {and} \bibinfo{person}{Taghi~M
  Khoshgoftaar}.} \bibinfo{year}{2009}\natexlab{}.
\newblock \showarticletitle{A survey of collaborative filtering techniques}.
\newblock \bibinfo{journal}{\emph{Advances in artificial intelligence}}
  \bibinfo{volume}{2009} (\bibinfo{year}{2009}), \bibinfo{pages}{4}.
\newblock


\bibitem[\protect\citeauthoryear{Sun, Han, Yan, Yu, and Wu}{Sun
  et~al\mbox{.}}{2011}]%
        {sun2011pathsim}
\bibfield{author}{\bibinfo{person}{Yizhou Sun}, \bibinfo{person}{Jiawei Han},
  \bibinfo{person}{Xifeng Yan}, \bibinfo{person}{Philip~S Yu}, {and}
  \bibinfo{person}{Tianyi Wu}.} \bibinfo{year}{2011}\natexlab{}.
\newblock \showarticletitle{Pathsim: Meta path-based top-k similarity search in
  heterogeneous information networks}.
\newblock \bibinfo{journal}{\emph{Proceedings of the VLDB Endowment}}
  \bibinfo{volume}{4}, \bibinfo{number}{11} (\bibinfo{year}{2011}),
  \bibinfo{pages}{992--1003}.
\newblock


\bibitem[\protect\citeauthoryear{Tang, Qu, Wang, Zhang, Yan, and Mei}{Tang
  et~al\mbox{.}}{2015}]%
        {tang2015line}
\bibfield{author}{\bibinfo{person}{Jian Tang}, \bibinfo{person}{Meng Qu},
  \bibinfo{person}{Mingzhe Wang}, \bibinfo{person}{Ming Zhang},
  \bibinfo{person}{Jun Yan}, {and} \bibinfo{person}{Qiaozhu Mei}.}
  \bibinfo{year}{2015}\natexlab{}.
\newblock \showarticletitle{Line: Large-scale information network embedding}.
  In \bibinfo{booktitle}{\emph{Proceedings of the 24th International Conference
  on World Wide Web}}. International World Wide Web Conferences Steering
  Committee, \bibinfo{pages}{1067--1077}.
\newblock


\bibitem[\protect\citeauthoryear{Tang and Zhang}{Tang and Zhang}{2009}]%
        {tang2009discriminative}
\bibfield{author}{\bibinfo{person}{Jie Tang} {and} \bibinfo{person}{Jing
  Zhang}.} \bibinfo{year}{2009}\natexlab{}.
\newblock \showarticletitle{A discriminative approach to topic-based citation
  recommendation}.
\newblock \bibinfo{journal}{\emph{Advances in Knowledge Discovery and Data
  Mining}} (\bibinfo{year}{2009}), \bibinfo{pages}{572--579}.
\newblock


\bibitem[\protect\citeauthoryear{Tang, Wan, and Zhang}{Tang
  et~al\mbox{.}}{2014}]%
        {tang2014cross}
\bibfield{author}{\bibinfo{person}{Xuewei Tang}, \bibinfo{person}{Xiaojun Wan},
  {and} \bibinfo{person}{Xun Zhang}.} \bibinfo{year}{2014}\natexlab{}.
\newblock \showarticletitle{Cross-language context-aware citation
  recommendation in scientific articles}. In
  \bibinfo{booktitle}{\emph{Proceedings of the 37th international ACM SIGIR
  conference on Research \& development in information retrieval}}. ACM,
  \bibinfo{pages}{817--826}.
\newblock


\bibitem[\protect\citeauthoryear{Zhai and Lafferty}{Zhai and Lafferty}{2001}]%
        {zhai2001study}
\bibfield{author}{\bibinfo{person}{Chengxiang Zhai} {and} \bibinfo{person}{John
  Lafferty}.} \bibinfo{year}{2001}\natexlab{}.
\newblock \showarticletitle{A study of smoothing methods for language models
  applied to ad hoc information retrieval}. In
  \bibinfo{booktitle}{\emph{Proceedings of the 24th annual international ACM
  SIGIR conference on Research and development in information retrieval}}. ACM,
  \bibinfo{pages}{334--342}.
\newblock


\end{thebibliography}
